\documentstyle[prb,aps,epsf]{revtex} \def\narrowtext{} \tighten \twocolumn
\input epsf.sty
\begin{document}
\draft
 
\title{Collective Modes and the Superconducting State Spectral Function
of Bi2212}
\author{M. R. Norman$^1$ and H. Ding$^{1,2}$}
\address{
         (1) Materials Science Division, Argonne National Laboratory,
             Argonne, IL 60439 \\
         (2) Department of Physics, University of Illinois at Chicago,
             Chicago, IL 60607\\
         }

\address{%
\begin{minipage}[t]{6.0in}
\begin{abstract}
Photoemission spectra of the high temperature
superconductor Bi2212 near $(\pi,0)$ show a dramatic change when 
cooling below $T_c$:  the broad peak in the normal state turns into a sharp
low energy peak followed by a higher binding energy hump.  Recent experiments
find that this low energy peak persists over a significant range in
momentum space.  We show in this paper that these data are well described by
a simple model of electrons interacting with a collective mode which appears
only below $T_c$.
\typeout{polish abstract}
\end{abstract}
\pacs{PACS numbers: 71.25.Hc, 74.25.Jb, 74.72.Hs, 79.60.Bm}
\end{minipage}}

\maketitle
\narrowtext

Angle resolved photoemission spectroscopy (ARPES) has become one of the key
tools used to elucidate the physics of high temperature superconductors.
It has produced a number of important observations concerning the nature of
the normal and superconducting phases.  Examples are the existence of a large
Fermi surface, and an anisotropic energy gap in the superconducting and
pseudogap phases\cite{REVIEW}.  The most interesting aspect of the
ARPES data, though, is the unusual nature of the spectral lineshape
and how this lineshape changes as a function of doping, momentum, and
temperature.  Perhaps the most profound example in this regard
is the temperature dependence of the lineshape near the $(\pi,0)$ point in
Bi2212. A very broad normal state spectrum evolves quite rapidly below
$T_c$ into a resolution limited quasiparticle peak,
followed at higher binding energies by a dip
then a hump, after which the spectrum is equivalent to that in the normal
state.\cite{DESSAU,NK,DING96}
Similar effects have been seen in tunneling spectra, where it has been
found that all of these spectral features (peak, dip, hump) scale with
the superconducting gap\cite{TUNNEL}.  This implies that the electron
self-energy has a dramatic change below $T_c$.

In a recent paper, our group has shown that the low energy peak persists
over a surprisingly large range in momentum space along the $(\pi,0)-(0,0)$
and $(\pi,0)-(\pi,\pi)$ directions\cite{NORM97}.  As argued in that paper,
this result can be connected to the change in lineshape with temperature
noted above.  The idea is that the dip in the spectrum at $(\pi,0)$ implies
that the imaginary part of the self-energy, $Im\Sigma$, has a step-like
drop from a large value at binding energies larger than the dip to a small
value for smaller energies.  This step behavior has recently been verified by
us by a direct extraction of $\Sigma$ from the data\cite{NEW}.  By
Kramers-Kronig transformation, then, $Re\Sigma$ will have a strong peak at the
dip energy.  The consequence of this is that there will always be a low energy
quasiparticle pole trapped on the lower binding energy side of the dip energy,
even when the normal state binding energy is quite large.  It is this effect
which we believe leads to the persistent peak.

This raises the question of what kind of microscopic picture can lead to such
behavior.  As discussed in our paper\cite{NORM97}, a step edge in $Im\Sigma$
is equivalent to the problem of an electron interacting with a sharp
(dispersionless) mode.  This model has been worked out in detail in
the classic literature of strong-coupling superconductors, where the mode is
an Einstein phonon\cite{OLD}.  In the current case, though, the effect of the
mode only appears below $T_c$, and therefore implies a collective mode of
electronic origin.  Still, the mathematics is largely equivalent.  What we
show in this paper is that this simple model gives a good quantitative fit to
the data.

\begin{figure}
\epsfxsize=2.4in
\epsfbox{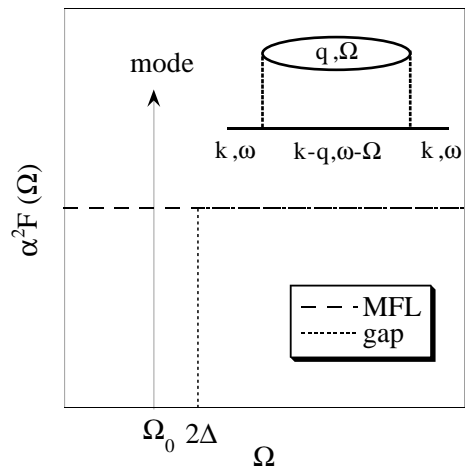}
\vspace{0.5cm}
\caption{
$\alpha^2F$ for three models, MFL 
(dashed line), gapped MFL (dotted line), and gapped MFL plus mode (dotted
line plus $\delta$ function).  Inset:
Feynman diagram for the lowest order contribution to $\Sigma$ from 
electron-electron scattering.}
\label{fig1}
\end{figure}

We begin by discussing self-energy effects in superconductors.
For now, we ignore the complication of momentum dependence.  
The lowest order contribution to electron-electron scattering is 
represented by the Feynman diagram shown in the inset of Fig.~1.  In the 
superconducting state, each internal line will be gapped by $\Delta$.  
This implies that the scattering will be suppressed for $|\omega| < 
3\Delta$\cite{KURODA}.  This explains the presence of a sharp 
resolution-limited quasiparticle peak at low temperatures.  What is not 
so obvious is whether this in addition explains the strong spectral dip.  
Explicit calculations show only a weak dip-like feature\cite{LITTLE}.  To 
understand this in detail, we equate the bubble plus interaction lines
(Fig.~1 inset) to an ``$\alpha^2F$'' as 
in standard strong-coupling literature\cite{OLD}.  In a marginal Fermi liquid 
(MFL) at T=0, $\alpha^2F(\Omega)$ is simply a constant in $\Omega$.  The 
effect of the gap is to force $\alpha^2F$ to zero for $\Omega < 
2\Delta$.  The question then arises where the gapped weight goes.  It 
could be distributed to higher energies, but in light of the above 
discussion, we might expect it to appear as a collective mode inside of 
the $2\Delta$ gap.  For instance, this indeed occurs in FLEX calculations where 
the bubble represents spin fluctuations\cite{PAO}, in which case a sharp 
mode will appear if the condition $1-U\chi_0({\bf q},\Omega)=0$ is 
satisfied for $\Omega < 2\Delta$.  These three cases (MFL, gapped MFL, 
gapped MFL plus mode) are illustrated in Fig.~1.

\begin{figure}
\epsfxsize=3.2in
\epsfbox{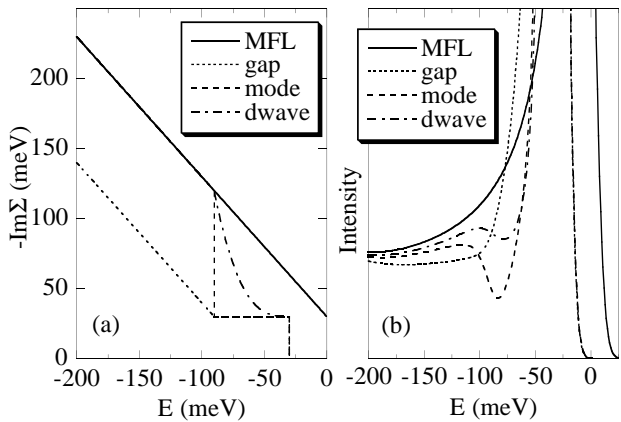}
\vspace{0.5cm}
\caption{
(a) $Im\Sigma$ for MFL (solid line), gapped MFL (dotted line), gapped 
MFL plus mode (dashed line), and simple d-wave model (dashed-dotted line).
Parameters are $\alpha$=1, $\omega_c$=200meV, $\Delta$=30meV (0 for MFL),
$\Omega_0$=2$\Delta$, and $\Gamma_0$=30meV.
(b) Spectral functions (times a Fermi function with $T$=14K convolved with a 
resolution gaussian of $\sigma$=7.5 meV) for these four cases
($\epsilon$=-34meV).}
\label{fig2}
\end{figure}

$\Sigma$ is easy to obtain analytically\cite{OLD} if we ignore the 
complication of the superconducting density of states from the 
$\bf k-q$ line of Fig.~1 and just replace this by a step function at 
$\Delta$.  The resulting $Im\Sigma$ for the gapped MFL and gapped MFL plus
mode models are shown in Fig.~2a in comparison to the normal state MFL. 
Note that structure in $\alpha^2F$ at $\Omega$ appears in $\Sigma$ at 
$|\omega|=\Omega+\Delta$ due to the gap in the $\bf k-q$ line.  Moreover, the
MFL plus mode is simply the normal state MFL cut-off at $3\Delta$ (this is
obtained under the assumption that all the gapped weight in $\alpha^2F$ 
shows up in the mode).  In contrast, the gapped MFL decays linearly to zero
at $3\Delta$.
This simple picture is not changed that much if one actually solves 
the strong-coupling equations for $\Sigma$ and 
$\Delta$\cite{LITTLE}.  (For an s-wave gap, the linear 
behavior of $Im\Sigma$ is replaced by a square root behavior).

The spectral function is given by\cite{OLD}
\begin{equation}
A(\omega) = \frac{1}{\pi}Im \frac{Z\omega + \epsilon}
{Z^2(\omega^2-\Delta^2)-\epsilon^2}
\end{equation}
with (a complex) $Z(\omega) = 1 - \Sigma(\omega)/\omega$.
These are shown in Fig.~2b and were convolved with a gaussian of
$\sigma$=7.5 meV, typical of high resolution ARPES, with a
constant $Im\Sigma$ ($\Gamma_0$) added
for $|\omega| > \Delta$ to reduce the size of the quasiparticle peak.
We note that there is no dip as such for the gapped MFL model, whereas the
addition of the mode causes a significant dip.  The latter behavior is
consistent with 
experiment.  Moreover, the mode model has the additional advantage that  
$Im\Sigma$ recovers back to the normal state value by $3\Delta$, which is 
also in agreement with experiment in that the normal and superconducting 
state spectra agree beyond 90 meV\cite{NORM97}.

We contrast this behavior with that expected for a simple d-wave 
model.  To a first approximation, this can be obtained by replacing the 
step drop in $Im\Sigma$ in the MFL plus mode model with 
$(|\omega|-\Delta)^3$ for $|\omega| < 3\Delta$\cite{QUINLAN}.  This is 
shown in Fig.~2a as well, with the resulting spectrum in Fig.~2b.  Only a
weak dip appears.  Moreover, we have analyzed models with 
the exponent 3 replaced by some n and have found that n must be large
to obtain a dip as strong as seen in experiment.  
Therefore, the upshot is that at the least, something similar to a step 
is required in $Im\Sigma$ to be consistent with experiment.

In principle, we could take the above MFL plus mode model and fit 
experiment with it.  In this paper, though, we consider a simpler model.  
There are several reasons for this.  First, the MFL model has a number 
of adjustable parameters associated with it.  There is the coupling
constant ($\alpha$), the cut-off frequency ($\omega_c$), and
the mode energy (which is not in general $2\Delta$).  Moreover, 
the spectrum for $\bf k$ points near the $(\pi,0)$ point does not appear to be
MFL-like in nature.  We have found that the 
normal state Bi2212 spectrum is fit very well by a Lorentzian plus a 
constant in an energy range less than $0.5$eV.  This is also true for
Bi2201 spectrum where 
the normal state can be accessed to much lower temperatures.  The 
constant term represents the so-called ``background'' contribution, and 
is essentially equivalent to spectrum for $k > k_F$, where it is also 
seen that the background gets gapped by $\Delta$ in the 
superconducting state.  There are several possibilities for what the 
background could be due to, and in fact could be a combination of all of 
these: (1) incoherent part of $A$, (2) inelastic secondaries, (3) emission 
from the BiO layers, (4) diffraction of the photoelectrons by the surface BiO 
layer, etc.  Since this has little to do with the peak/dip/hump 
structure, we choose to subtract this off, but note the caveat that this 
is an incomplete description if part of the background is intrinsic.
Finally, the Lorentzian simplification allows us to directly obtain the
dispersion $\epsilon_{\bf k}$ from tight binding fits to  
the normal state peak positions\cite{DING96}.

In the resulting Lorentzian model, the normal state $\Sigma$ is purely an 
imaginary constant, and $\alpha^2F$ is a mode at zero 
energy.  In the superconducting state, this mode gets pushed back to some 
energy within $2\Delta$.  This model is artificial in the sense that 
all the self-energy is being generated by the mode.  That is why we went 
through the above discussion motivating the mode more properly as a 
rearrangement of $\alpha^2F$ due to the superconducting gap.  In 
practice, though, the results are very similar to the MFL plus mode model, 
and has the further advantage of having the several parameters of that 
model collapse to just the mode strength ($\Gamma_1$) and mode position 
($\Omega_0$) of the Lorentzian model.  Moreover, analytic results can 
still be obtained for $\Sigma$ when the superconducting density of states 
for the $\bf k-q$ line of Fig.~1 is taken into account.  The 
result is
\begin{eqnarray}
-Im\Sigma(\omega)&=&\Gamma_0N(|\omega|) + \Gamma_1N(|\omega|-\Omega_0), \;
 |\omega| > \Omega_0+\Delta \nonumber \\
                &=&\Gamma_0N(|\omega|), \;
 \Delta < |\omega| < \Omega_0+\Delta \nonumber \\
                &=&0, \;  |\omega| < \Delta
\end{eqnarray}
where $N(\omega)= \omega/\sqrt{\omega^2-\Delta^2}$ 
is the BCS density of states, and
\begin{eqnarray}
\pi Re\Sigma(\omega) = \Gamma_0N(-\omega) \ln\left[{|-\omega+
\sqrt{\omega^2-\Delta^2}|}/{\Delta}\right] \nonumber \\
 + \Gamma_1N(\Omega_0-\omega) \ln\left[{|\Omega_0-\omega
+\sqrt{(\omega-\Omega_0)^2-\Delta^2}|}/{\Delta}\right] \nonumber \\
- \{\omega \rightarrow -\omega\}
\end{eqnarray}
where it has again been assumed that $\Delta$ is a real constant in 
frequency.  An s-wave density of states has been used to obtain an 
analytic result.  A d-wave density of states will not be that different.
The advantage of 
an analytic result is that it is useful when having to take spectra and 
convolve with resolution to compare to experiment. 
Our results are not very sensitive to $\Gamma_0$ (30 meV), included 
again to damp the quasiparticle peak.  (A more realistic
damping of the peak would require making $\Delta$ complex.)
We use the same set of 
parameters for all $\bf k$ ($\Gamma_1$=200 meV), and therefore
assume a d-wave gap $\Delta_{\bf k} = \Delta_{max}(\cos(k_xa)-\cos(k_ya))/2$ 
in Eqs.~1-3 with $\Delta_{max}$ = 32 meV.
The best agreement with experiment is found by choosing
the mode energy $\Omega_0 = 1.3\Delta_{\bf k}$, so that the 
spectral dip for $(\pi,0)$ is at $2.3\Delta_{max}$.

The resulting real and imaginary parts of $\Sigma$ at $(\pi,0)$ are shown
in Fig.~3a. Note the singular behaviors 
at $\Delta$ due to the $\Gamma_0$
term and at $\Omega_0+\Delta$ due to the $\Gamma_1$ term.  In both cases, 
step drops in $Im\Sigma$ would also give singularities in $Re\Sigma$.  
The advantage of peaks in $Im\Sigma$ (due to the density of states)
is that it makes the dip deeper in better agreement with experiment.
In Fig.~3b and 3c, we show a comparison of the resulting spectral
function (convolved with the experimental energy and momentum resolution)
to experimental data at $(\pi,0)$ for both wide and narrow energy
scans, where a step edge background with a gap of $\Delta$ is added
to the calculated spectrum as discussed above.  The resulting agreement 
is excellent.

\begin{figure}
\epsfxsize=3.2in
\epsfbox{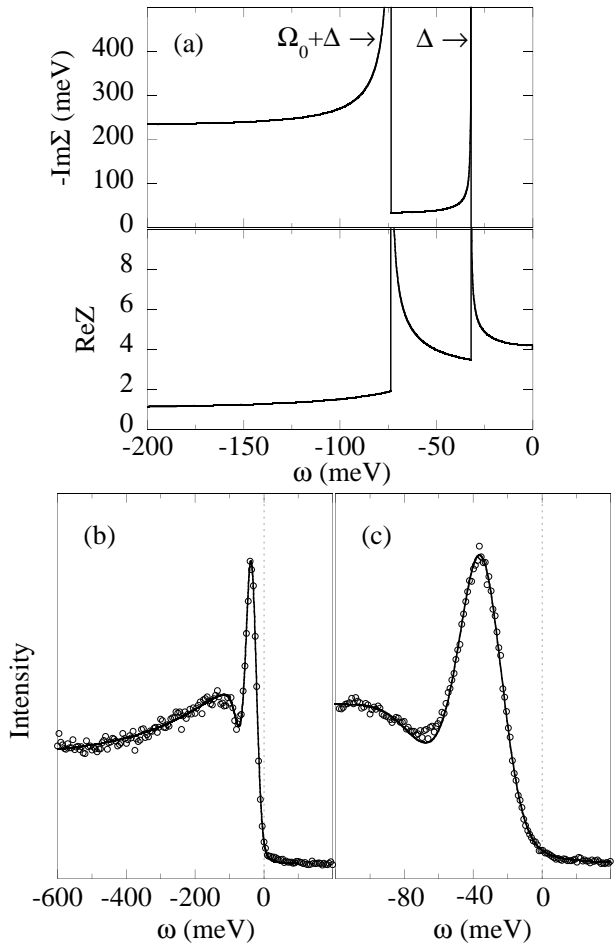}
\vspace{0.5cm}
\caption{
(a) $Im\Sigma$ and $ReZ$ at $(\pi,0)$ from Eqs.~2-3
($\Gamma_1$=200meV, $\Gamma_0$=30meV, $\Delta$=32meV, $\Omega_0$=1.3$\Delta$).
Comparison of the data at $(\pi,0)$ for (b) wide and (c) narrow energy scans
with calculations based on Eqs.~1-3, with an added 
step edge background contribution.}
\label{fig3}
\end{figure}

To better appreciate these results,
the positions of the sharp peak and the higher binding energy hump obtained
from the calculations are plotted relative to the normal state binding energy 
$\epsilon_{\bf k}$ along the $(0,0)-(\pi,0)$ direction, and compared to
those obtained from the experimental data of Ref.~\onlinecite{NORM97}
in Fig.~4.  
This plot is very similar to that obtained for electrons interacting with 
an Einstein mode\cite{OLD}.  The calculations reproduce
the dispersionless nature of the low frequency peak, as well as its lack
of visibility for $\bf k$ vectors close enough to $(0,0)$.  The dispersionless
behavior is due to several factors: 
(1) the weakness of the dispersion $\epsilon_{\bf k}$
near $(\pi,0)$, (2) the lowering of $\Omega_0 \sim \Delta_{\bf k}$ 
as one moves towards $(0,0)$, and (3) the influence of both
$Re\Sigma$ and $\Delta$.  The last is a new effect worth commenting on.
The real part of the self-energy implies a mass enhancement ($Z > 1$)
in the superconducting state relative to the normal state, which acts to
push spectral weight towards $E_F$.  On the other hand, $\Delta$
itself pushes spectral weight away from $E_F$.  Thus the
dispersion is dramatically flattened relative to the normal state.

\begin{figure}
\epsfxsize=2.4in
\epsfbox{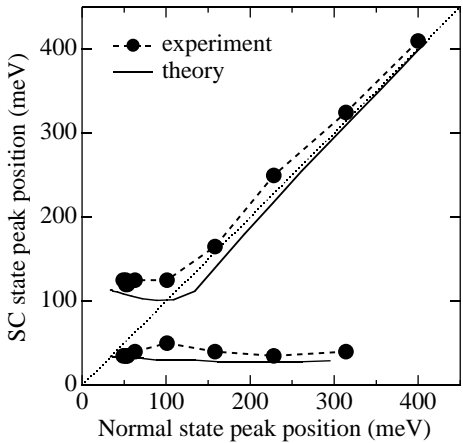}
\vspace{0.5cm}
\caption{
Positions (meV) of the sharp peak and the broad hump in the 
superconducting state versus normal state peak position.  
Solid points connected by a dashed line are the experimental data, 
the solid lines are obtained from the calculations,
and the dotted line represents the normal state dispersion.}
\label{fig4}
\end{figure}

In the above calculations, it was assumed that the mode frequency was 
proportional to $\Delta_{\bf k}$.  This was the easiest way we found to 
properly simulate the loss of the experimental low frequency peak as one 
disperses towards $(0,0)$.  In reality, $\Omega_0$ is a function of $\bf q$
in the diagram of Fig.~1, not $\bf k$. 
Moreover, it was our assumption of independence of $\Omega_0$ on 
$\bf q$ that allowed us to obtain a step drop in $Im\Sigma$,
leading to the spectral dip. 
Without some microscopic theory, only qualitative
observations can be made at this stage concerning the true dependence
of $\Sigma$ on momentum $\bf k$\cite{SHEN}.
Assuming an artificial limit where only ${\bf q}={\bf Q}=(\pi,\pi)$
contributes, we would replace $\Gamma_1N(\omega+\Omega_0)$ in Eq.~2 by 
$g_{k,k+Q}^2 A_{k+Q}(\omega+\Omega_0)$ (for $\omega<0$)
where $g$ is the interaction vertex.
Using a quasiparticle pole approximation for $A$ when solving
Eqs.~2-3, this would imply a dip in the spectrum
at $|\omega|=E_{k+Q}+\Omega_0$ where $E_k^2=\epsilon_k^2/(Re Z_k)^2
+\Delta_k^2$, and a persistent low frequency peak if $Z$ is large
enough.  The coupling of $\bf k$ and $\bf k+Q$ in the self-energy equations
also implies that if a low frequency peak exists for $\bf k$, then one also
exists for $\bf k+Q$.  This is just the effect observed in the data along
$(\pi,0)$-$(\pi,\pi)$\cite{NORM97}, in that a
low frequency peak exists for about the same
momentum range as that along $(\pi,0)$-$(0,0)$.  It remains to be seen
whether such simple momentum dependent models give 
as good a fit to the spectra as the dispersionless model presented here.

Finally, it is interesting to note that the mode energy we infer from 
the data is 41meV, equivalent (probably fortuitously) to a resonant mode 
energy observed in YBCO by neutron scattering data\cite{YBCO} at ${\bf 
Q}=(\pi,\pi)$.  The models proposed for this mode are similar to the model
discussed in this paper\cite{YTHEORY}.
So far, neutron scattering data on Bi2212 have yet to see a 
similar structure\cite{MOOK}, although these experiments were done on a rod
of aligned small crystals.  Our results here would imply that such 
experiments on large single crystals would be of interest. 

In conclusion, we have shown that a simple model of an electron 
interacting with a collective mode in the superconducting state gives 
a quantitative description of the unusual spectral lineshape seen by 
ARPES data in the superconducting state of Bi2212.  This implies that 
electron-electron scattering plays a dominant role in high temperature 
superconductors, and is in support of an electron-electron origin for the 
pairing.

We thank J.C. Campuzano and Mohit Randeria for many discussions on these
issues.  This work was supported by the U. S. Dept. of Energy,
Basic Energy Sciences, under contract W-31-109-ENG-38, the National 
Science Foundation DMR 9624048, and
DMR 91-20000 through the Science and Technology Center for
Superconductivity.

\end{document}